\documentclass[reprint, amsmath,amssymb, aps, pra]{revtex4-2}
\usepackage[table]{xcolor}
\usepackage{booktabs}
\usepackage{array}
\usepackage{wrapfig}
\usepackage{bm,physics,amsmath,amssymb,color}
\usepackage{ragged2e}
\usepackage[utf8]{inputenc}
\usepackage[top=0.6in, bottom=1in, left=0.8in, right=0.8in]{geometry}
\usepackage[normalem]{ulem}
\usepackage{graphicx,layouts}

\begin{document}

\title{Alignment-free ultra-broadband parametric frequency conversion in\\ lead-halide perovskites}

\author{Abhishek Shiva Kumar$^1$}
\author{Dusan Lorenc$^{1,2,3}$}
\author{Ayan A. Zhumekenov$^{4,5}$}
\author{Osman M. Bakr$^{4,6}$}
\author{Zhanybek Alpichshev$^1$}
\email{alpishev@ist.ac.at}

\affiliation{$^1$Institute of Science and Technology Austria (ISTA), Am Campus 1, 3400 Klosterneuburg, Austria}
\affiliation{$^2$International Laser Centre, Ilkovičova 3, 84104 Bratislava, Slovakia}
\affiliation{$^3$RCQI, Institute of Physics SAS, Dúbravská 9, 84104 Bratislava, Slovakia}
\affiliation{$^4$King Abdullah University of Science and Technology, Thuwal, Kingdom of Saudi Arabia}
\affiliation{$^5$School of Materials Science and Engineering (MSE), Nanyang Technological
University, Singapore, Singapore}
\affiliation{$^6$Center for Renewable Energy and Storage Technologies (CREST), Division of Physical Science and Engineering (PSE), King Abdullah University of Science and Technology, Thuwal, Kingdom of Saudi Arabia}

\begin{abstract}
Lead-halide perovskites(LHPs) were demonstrated to exhibit some of the largest known optical nonlinearities, yet their potential for frequency conversion remains largely untapped. 
Here we demonstrate ultrabroadband four-wave mixing of near- and mid-infrared femtosecond pulses in thick single-crystal LHPs, generating bright, coherent, and highly collimated emission across an exceptionally wide continuous tuning range without phase-matching engineering, angular alignment, or dispersion optimization. Time resolved measurements reveal that the emission originates near the crystal surfaces, where phase-matching constraints are relaxed, while the unusually large intrinsic $\chi^{(3)}$ response preserves efficient and directional frequency conversion despite the strongly localized interaction volume. These results position LHPs as a powerful bulk platform for ultrabroadband nonlinear photonics, opening a pathway toward compact, alignment-free architectures for ultrafast frequency conversion.
\end{abstract}

\maketitle
\noindent \textbf{Introduction.} Over the past decade, lead-halide perovskites (LHPs) have emerged as a promising class of materials owing to their outstanding optoelectronic properties, which have enabled applications in light-emitting diodes \cite{Lin2018,Du2021,Kim2022}, solar cells \cite{Lin2022,Chen2023,Dai2022}, and lasers \cite{Zhang2014,Qin2020,Sun2020,Han2020,Wang2023}. Beyond these applications, LHPs exhibit a range of unusual optical phenomena, including superradiance \cite{Biliroglu2025}, bright triplet excitons \cite{Becker2018}, and laser cooling \cite{Ha2016}, as well as exceptionally large third-order nonlinearities \cite{ma15010389}. In particular, lead bromide perovskites exhibit nonlinear refractive index $n_2$ values 
that place them among the most nonlinear bulk semiconductors reported to date \cite{Yi17,Zhang2016,Kalanoor2016,Krishnakanth18,10.1063/1.5090926,MIRERSHADI20167,NADAFAN2023109055}. These properties make LHPs attractive candidates for nonlinear frequency conversion, although their capabilities in this context remain largely unexplored.

Four Wave Mixing (FWM), the nonlinear interaction of three optical fields generating a fourth, constitutes one of the most versatile nonlinear optical processes, which has been extensively explored across modern photonics for applications in communications \cite{1016354}, computing \cite{Gu:24}, metrology \cite{Hudelist2014}, imaging \cite{Lemos2014} and quantum optics \cite{Wang2018}, enabling a wide range of functionalities spanning wavelength conversion \cite{JianhuiZhou1994, Melloni2008, Noskovicova2024}, frequency comb generation \cite{Dong2016, photonics9100746, Yang:16, Wang2013, Husakou2003, Xue2017}, optical amplification \cite{Stolen1982,Kuyken11}, all-optical switching \cite{Diez1997,Pscherer2021}, signal regeneration \cite{Salem2008}, logic operations \cite{Li11}, quantum memory construction \cite{Radnaev2010}, and the generation of entangled photon pairs \cite{Pooser:09, Takesue2004,Harada:08, PhysRevA.93.033810, Fang:13,PhysRevA.107.013514, Zheltikov2017, Srivathsan2013, PhysRevLett.114.063902}.  Despite its broad applicability, achieving efficient, widely tunable, and ultrabroadband FWM has proven challenging in practice. Conventional approaches rely on precise phase matching, non-collinear geometries, or cascaded schemes requiring careful control of alignment, dispersion, and temporal synchronization \cite{Harrison1979,Liu:08, Liu:09, Zhang:11, Crespo:00, Lu:14, He:13, Liu:13, s100504296, Wang2014, Zhu2013}. To overcome these limitations, a wide range of engineered platforms have been explored, including near-zero-index materials \cite{Carnemolla:21, Suchowski2013} and nanophotonic implementations such as waveguides \cite{Kowligy2018, Zhao2025, Timmerkamp2025}, plasmonic systems \cite{Dai2021}, and optical fibers \cite{Corso2023}. While these approaches can enhance nonlinear interactions or relax phase-matching constraints, they typically do so at the cost of increased complexity, reduced pulse energies, or limited bandwidth. A simple, alignment-insensitive platform for broadband nonlinear frequency conversion therefore remains elusive.

\begin{figure*}[t]
    \centering
    \includegraphics[width=\linewidth]{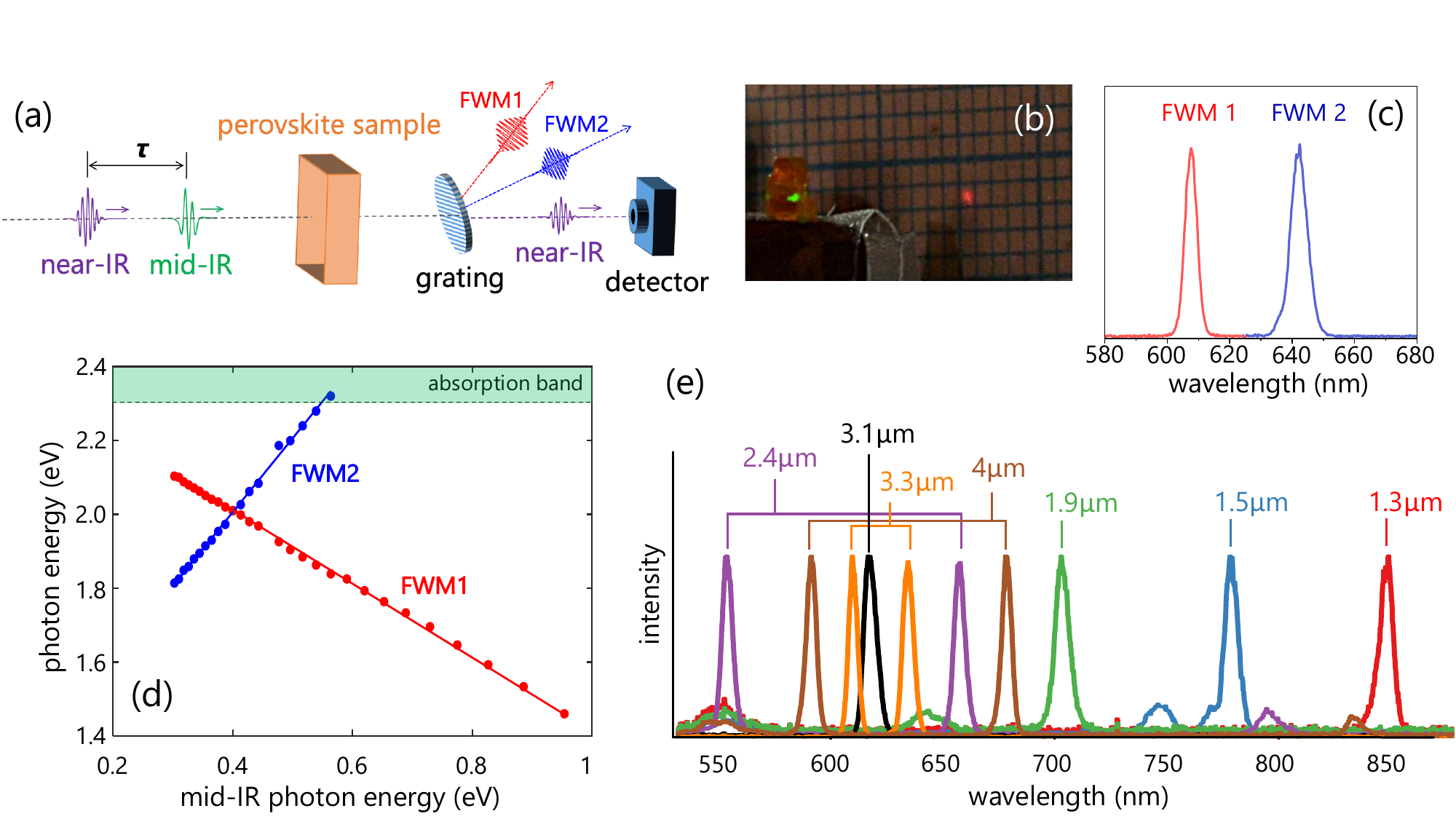}
    \caption{
    \textbf{Ultra-broadband FWM emission from single-crystal perovskite.}
(a) Schematic of our experimental configuration. $\tau$ denotes the time delay between the near-IR and mid-IR pulses.
(b) Photograph of FWM emission from a 1-mm-thick CsPbBr$_3$ single crystal, projected onto millimeter-scale graph paper positioned 1.3 cm behind the sample. The green surface glow arises from photoluminescence induced by multi-photon absorption of pump pulses.
(c) Representative FWM spectrum for MIR wavelength 3.4 $\mu$m, showing two distinct components: $2\omega_{\mathrm{NIR}} - \omega_{\mathrm{MIR}}$ (FWM1) and $\omega_{\mathrm{NIR}} + 2\omega_{\mathrm{MIR}}$ (FWM2).
(d) Photon energies of FWM1 (red dots) and FWM2 (blue dots) as a function of mid-IR photon energy. The measured energies follow the relations $2\omega_{\mathrm{NIR}} - \omega_{\mathrm{MIR}}$ (red line) and $\omega_{\mathrm{NIR}} + 2\omega_{\mathrm{MIR}}$ (blue line). 
(e) Representative FWM spectra for selected mid-IR excitation wavelengths in the range 1.3 – 4$\mu$m}
    \label{fig:broadband}
\end{figure*}

Here, we show that lead-halide perovskites exhibit strong   four-wave mixing characterized by ultrabroadband pulse-matching-free frequency conversion in a simple bulk geometry. By mixing collinear near-infrared (near-IR) and mid-infrared (mid-IR) femtosecond pulses, we observe intense, highly collimated emission spanning a very wide spectral range without the need for phase-matching engineering, angular alignment, or
dispersion optimization. Notably, the generated FWM signal is sufficiently strong to be directly visible to the naked eye, which is highly unusual, particularly under conditions of strong phase mismatch. Fluence- and polarization-dependent measurements establish the coherent parametric $\chi^{(3)}$ nature of the process, while delay-resolved experiments reveal that the nonlinear interaction is localized near the sample surface. This surface-limited response relaxes conventional phase-matching constraints while the exceptionally high intrinsic $\chi^{(3)}$ of LHP keeps the emission bright and collimated, and enables efficient broadband conversion despite strong dispersion mismatch. Together, these results identify lead-halide perovskites as a powerful platform for broadband nonlinear photonics and point toward compact, alignment-insensitive architectures for ultrafast frequency conversion.

\noindent {\bf Experimental Results.} In this work we focus on two representative perovskite compounds: CsPbBr$_3$ and MAPbBr$_3$. Both can be grown into large high-quality single-crystal samples, thus avoiding the complexities of separating the intrinsic material response from that of structural defects such as grain boundaries in polycrystalline films and finite size effects in nanocrystals. Moreover, the two systems share the same inorganic cage and therefore have very similar optical properties such as bandgap ($\Delta \approx 2.3$eV \cite{Saidaminov2015, Saidaminov2016}) and refractive index \cite{Ishteev2022, Brennan2024}.  As a result, the observations on both can be treated as complementary to each other, with noticeable differences appearing only when material-specific effects become significant. Notably, MAPbBr$_3$ has cubic structure above $T_{M}$=240K, but has a series of distinct absorption bands in mid-infrared range due to vibration modes of methylammonium cation. In contrast, CsPbBr$_3$ is transparent below interband absorption edge ($\hbar \omega <\Delta$), but needs to be heated above $T_{C}=$403K to enter the optically isotropic cubic phase.

\begin{figure*}[t]
    \centering
    \includegraphics[width=\linewidth]{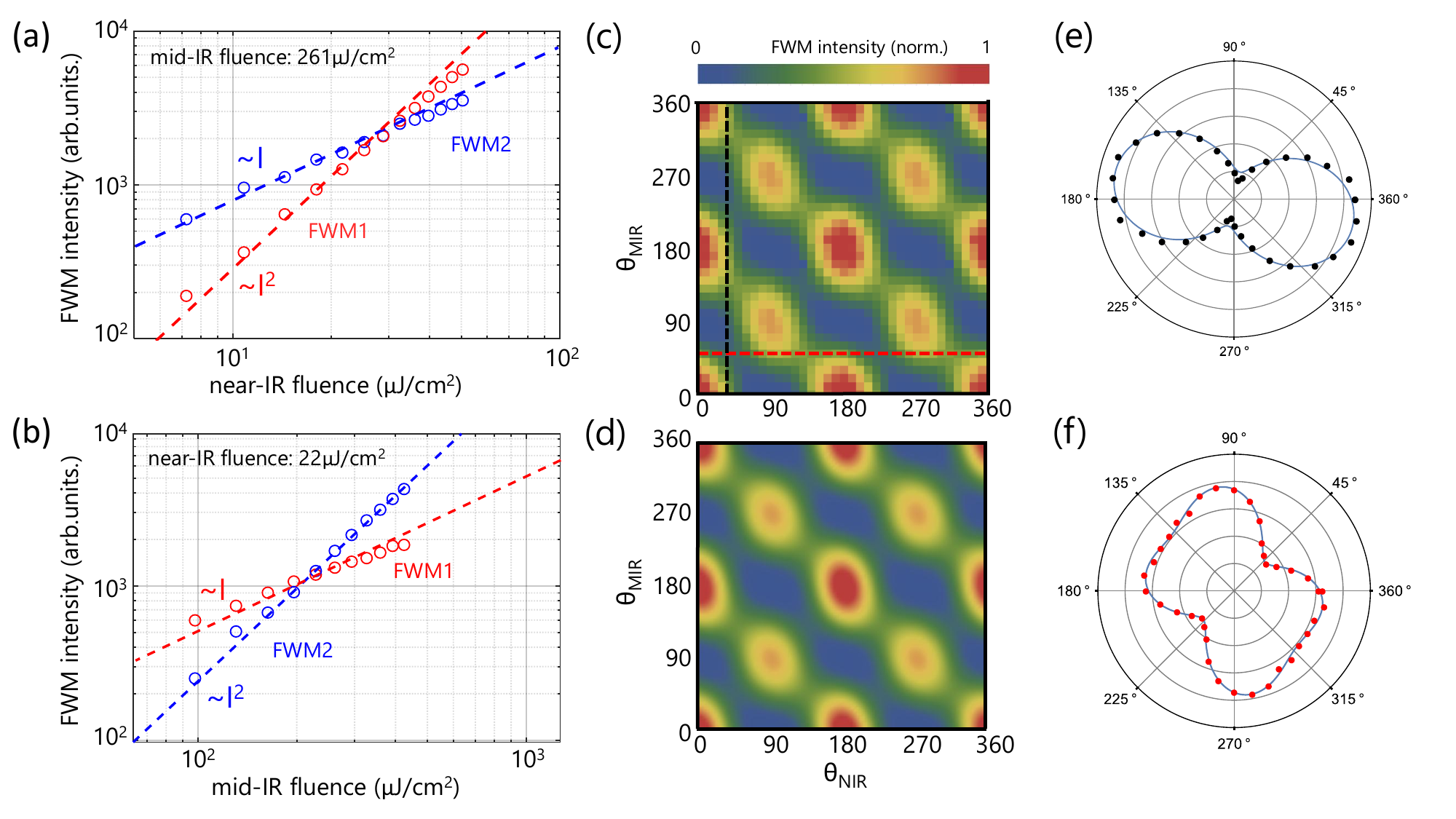}
    \caption{
\textbf{Parametric nature of FWM.}
(a,b) Intensities of FWM components ($I_{\mathrm{FWM1}}$ and $I_{\mathrm{FWM2}}$) as a function of (a) near-IR- ($I_{\mathrm{NIR}}$) and (b) mid-IR fluence ($I_{\mathrm{MIR}}$) consistent with expected intensity scaling (see text);
(c) Experimental dependence of $I_{\mathrm{FWM1}}$ on mid-IR- ($\theta_{\mathrm{MIR}}$) and near-IR ($\theta_{\mathrm{NIR}}$) polarization relative to (cubic) crystal axes of the sample. The apparent lack of $90^{\circ}$-rotation symmetry in the pattern is due to polarization-dependent efficiency of the monochromator used;
(d) Theoretical fit for data in (c) using third-order $\chi^{(3)}$ tensor for cubic medium;
(e) Polar plot of FWM intensity versus $\theta_{\mathrm{MIR}}$ at fixed $\theta_{\mathrm{NIR}} = 30^\circ$;
(f) Polar plot of FWM intensity versus $\theta_{\mathrm{NIR}}$ at fixed $\theta_{\mathrm{MIR}} = 50^\circ$.
Together, the fluence- and polarization-dependent measurements unambiguously establish the coherent $\chi^{(3)}$ nature of the FWM response.}
    \label{fig:R vs barrier height}
\end{figure*}

\vspace{0.5em}

We induce nonlinear response in bulk single-crystal perovskite samples by mixing two pulsed laser beams with distinct wavelengths in a collinear geometry followed by spectral analysis of the output (Fig.\ref{fig:broadband}a). We keep the wavelengths of both beams below the absorption edge of the sample, to access the entire bulk and to minimize the interference from large population of photocarriers. One beam was kept fixed at $\hbar \omega_{\mathrm{NIR}} = 1.2$eV (1030~nm) and the other one tuned across a broad mid-infrared range  $0.30\textrm{eV}<\hbar \omega_{\mathrm{MIR}} < 0.95\textrm{eV}$ (1300–4100~nm). 

When exposed to radiation, both perovskite compounds can be observed to emit the characteristic photoluminescence (PL) at the wavelength corresponding to the bandgap of the material (see green glow in Fig.\ref{fig:broadband}b). This happens due to nonlinear multi-photon absorption and is well-documented in literature \cite{Lorenc2025}. The unusual aspect of the present setting however, is the fact that in addition to the standard diffuse PL, there is also a collimated beam radiating out from the sample. It is intense and is readily visible to a naked eye. It is collinear with both pump beams, and its wavelength can be tuned by changing their wavelengths (red spot on the screen in Fig.\ref{fig:broadband}b) indicating intrinsically ultra-broadband nonlinear frequency conversion. The spectral analysis of this novel emission resolves two distinct nonlinear mixing components (Fig.\ref{fig:broadband}c) whose frequencies are found to follow the relations $\Omega_1 = 2\omega_{\mathrm{NIR}}-\omega_{\mathrm{MIR}}$ (FWM1) and $\Omega_2 = \omega_{\mathrm{NIR}}+2\omega_{\mathrm{MIR}}$ (FWM2) as the mid-IR wavelength is tuned (Fig.\ref{fig:broadband}d; also see Supplementary Materials for MAPbBr$_3$ data), identifying the process as four-wave mixing. This is corroborated by the scaling of each peak with respect to pumping intensities: $I_{\mathrm{FWM1}}\propto I_{\mathrm{NIR}}^{2}I_{\mathrm{MIR}}$ and $I_{\mathrm{FWM2}}\propto I_{\mathrm{NIR}}I_{\mathrm{MIR}}^{2}$ for FWM1 and FWM2 respectively (Figs.\ref{fig:R vs barrier height}a,b). 

While the observed intensity scaling of the nonlinear frequency conversion is highly suggestive, it may, in principle, still be misleading. The reason is that, in addition to the intrinsic instantaneous coupling, the pump intensities can also affect the efficiency of a nonlinear process extrinsically, for example via parasitic heating of the sample. Such effects can obscure the true power scaling. To address this, we perform an independent test of the parametric third-order nonlinear character of the observed four-wave mixing by probing its dependence on the polarization of the input beams relative to the crystalline axes of the sample, at fixed pump intensities $I_{\mathrm{NIR}}$ and $I_{\mathrm{MIR}}$. As shown in Figs.\ref{fig:R vs barrier height}c–f, the observed polarization dependence exhibits a pattern fully consistent with the tensorial structure of the third-order susceptibility $\chi^{(3)}_{\alpha \beta \gamma \delta}$ in a cubic crystal. Note that for polarization-sensitive measurements on bulk samples, it is critical to work with optically isotropic materials for the polarizations of the beams involved to be well-defined. Here in Figs.\ref{fig:R vs barrier height}c-f we show the data obtained on MAPbBr$_3$ at room temperature; see Supplementary Materials for analogous data obtained on CsPbBr$_3$ heated above 400K into its cubic phase.

\begin{figure*}[t]
  \center
    \includegraphics[width=1.0\linewidth]{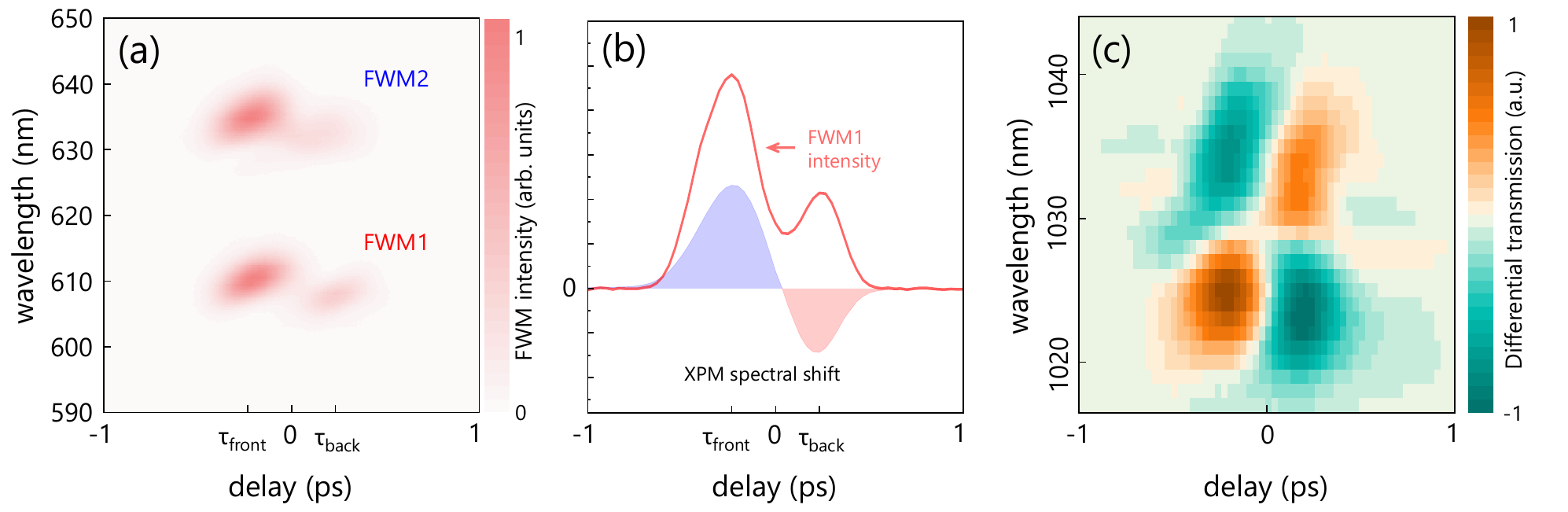}
    \caption{
    \textbf{Four-wave mixing vs. cross-phase modulation}
(a) FWM spectrum as a function of the relative delay between the near-IR and mid-IR pulses $\tau$ in CsPbBr$_3$.
(b) Integrated intensity of FWM1 and XPM spectral shift of the transmitted near-IR beam as function of $\tau$.
(c) Delay- and wavelength-resolved differential transmission of the near-IR beam in CsPbBr$_3$, revealing the characteristic pattern associated with XPM \cite{Lorenc2023}.}
    \label{fig:time trace}
\end{figure*}

Taken together, the evidence indicates that the novel spectral components are generated as a result of a canonical third-order-nonlinear parametric four-wave mixing process. The surprising aspect of the observation, though, is the fact, that based on reported refractive-index dispersion data for both CsPbBr$_3$ and MAPbBr$_3$, it is evident that the four-wave mixing in Fig.\ref{fig:R vs barrier height} is vastly out of phase-matching, with a coherence length $l_{\mathrm{coh}}\sim5\mu$m, much smaller than sample thickness $L\approx1$mm. 

In order to gain better understanding of the mechanism behind FWM in perovskites, we study the efficiency of this process as a function of delay time $\tau$ between mid-IR and near-IR pump pulses. As can be seen in Fig.\ref{fig:time trace}a FWM only occurs when the near- and mid-IR pulses overlap in time, as expected for an off-resonant process ($\hbar \Omega_{\textrm{FWM}}<\Delta$). Unexpectedly, instead of the standard bell-shaped $\tau$-dependence with a single maximum, the data indicates that the intensity of both FWM1 and FMW2 exhibit two distinct maxima. 

To clarify the origin of the unusual $\tau$-dependence, and shed light on the mechanism underlying the non-phase matched broadband frequency mixing in perovskites we consider cross-phase modulation (XPM): another $\chi^{(3)}$ process requiring phase-matching but whose mechanism is very well understood \cite{Lorenc2023}. In XPM, the time-dependent intensity of one optical field transiently modifies the refractive index experienced by the other, resulting in a spectral shift of the participating photons \cite{Boyd2000}. Since XPM is closely related to Kerr effect, a process phase-matched automatically, the fact that XPM  requires phase matching is often overlooked. The crucial difference is that unlike Kerr effect, XPM involves frequency shift $\delta \omega$ of the constituent fields between spectral modes at $(\omega,\Omega)$ and $(\omega-\delta\omega,\Omega+\delta\omega)$, where $\delta\omega$ is the XPM induced spectral shift in photon frequencies $\omega$ and $\Omega$. The condition for phase matching, is that total momentum of all involved photons is conserved:

\begin{equation*}
    k(\omega)+k(\Omega)=k(\omega-\delta\omega)+k(\Omega+\delta\omega)
\end{equation*}

\noindent when $\delta \omega \ll \omega,\Omega$, this is equivalent to the requirement that group velocity $v_g=\partial k/\partial \omega$ of both beams are equal:

\begin{equation*}
v_g(\omega) = v_g(\Omega)
\end{equation*}

 In a dispersive medium this condition is not satisfied in general. Therefore, naively, cross-phase modulation should not occur in uniform extended media. The reason why cross-phase modulation is observed in practice is because real samples have finite size. Specifically, in regions where strong gradients of optical properties break translational invariance, the momentum-conservation is relaxed, allowing XPM between non-degenerate pulses to occur \textit{e.g.} when the two pulses overlap spatio-temporally near sample surfaces \cite{Lorenc2023}.

Because of their markedly different group velocities, the specific location of spatio-temporal overlap of the two pulses within the sample can be controlled by the relative delay $\tau$ between the pulses (Fig. \ref{fig:time trace}c). When we measure XPM-induced spectral shift of the NIR probe beam $\delta \omega$ as a function of delay $\tau$ we observe in Fig. \ref{fig:time trace}b a recognizable pattern with the magnitude of XPM signal $|\delta \omega|$ being maximal at two delays values $\tau_{\mathrm{front}}$ and $\tau_{\mathrm{back}}$  that correspond to the configuration when the MIR pump and NIR probe pulses overlap at the front- and back surfaces of the sample, respectively (\cite{Lorenc2023}; also Supplementary Material). When we compare this with FWM (red curve in Fig. \ref{fig:time trace}b), we observe that it is also maximal at the same delay points $\tau_{\mathrm{front}}$ and $\tau_{\mathrm{back}}$. This observation indicates unequivocally that even in thick samples FWM is generated only near sample surfaces, thus providing a natural explanation for the lack of phase-matching requirement. It also naturally accounts for the emission of frequencies, which would otherwise be expected to be absorbed by the sample medium (Fig.~\ref{fig:broadband}d and Supplementary Fig.~S1).  
\vspace{0.2cm}

\noindent{\bf Discussion.} In bulk media, nonlinear frequency conversion is typically constrained by phase-matching requirements \cite{Boyd2000}. Ultrabroadband and widely tunable frequency conversion is realized here with remarkable simplicity due to strong third-order nonlinearity of perovskites. Although the spatial localization of FWM near sample surfaces is unambiguously established above, it hinges on the exceptionally high values on nonlinear susceptibility, that allows appreciable conversion efficiency even when effective interaction volume is limited to the immediate vicinity of the surface ($z<l_{\mathrm{coh}}$). In this light, it is in place to consider possible alternative FWM mechanisms that circumvent the phase mismatch and benefit from the entire bulk.

One possible mechanism for mitigating phase mismatch is an intensity-dependent modification of the effective refractive index via the Kerr effect, which could compensate for the intrinsic mismatch \cite{Ren2025}. However, such a scenario would be accompanied by a highly nontrivial dependence of the four-wave-mixing signal on pump intensities, while the measured signals follow the expected cubic scaling laws

Another possibility is that the missing momentum is transferred to the crystal as a whole, either mediated by static structural disorder (\textit{i.e.} random quasi-phase matching  \cite{Baudrier-Raybaut2004, Skipetrov2004, Savo2020,PhysRevLett.97.013902, Bravo-Abad:10,Ru:17}); or via dynamic lattice deformations (\textit{i.e.} phonon-assisted frequency conversion \cite{PhysRevApplied.22.014066, Isaienko2016}). However, both these scenarios are unlikely in a cubic or nearly cubic system such as lead-halide perovskites. Indeed, in order for the structural imperfections to be effective in relaxing momentum conservation, there have to be significant structural variations on the lengthscale comparable to $l_{\mathrm{coh}}$ along the propagation axis of the beams. However, since the overall crystal structure is (approximately) cubic on average, one should expect similar structural variations in the transverse direction as well, that would in general affect the collimation of the generated FWM beam. Given that in the experiment the diameter of the interaction region, determined by the spot sizes of near-IR and mid-IR pump beams, $D\sim$100$\mu$m$\gg l_{\textrm{coh}}$ such disorder-caused processes would produce a strongly dispersed output, in contrast to the highly collimated FWM emission observed in the experiment (see Fig.\ref{fig:broadband} and Fig.~S5). Incidentally, the collimated nature of FWM emission also rules out the possibility of enhancement of nonlinear susceptibility via localized field \cite{Ko2021} which is inconsistent with the observed collimated nature of the emitted FWM beam. 

Another intriguing possibility is raised by the recent reports of cooperative emission phenomena in LHPs \cite{Findik2021, Biliroglu2022}. Namely, that the effective self-phase matching can be mediated by collective coherence. However, such mechanisms would require characteristic coherence-build-up dynamics~\cite{Kobiyama2026}, while our delay scans reveal an essentially instantaneous response.

Taken together, the available evidence indicates that the observed FWM in perovskites occurs via standard non-phase matched third-order nonlinear interaction of the beams confined to the vicinity of the surface($\lesssim 5\mu$m). This, though, is contingent upon correspondingly large nonlinear susceptibility to account for the observed bright emission. It must be mentioned here that while extremely high values of nonlinear susceptibilities have been reported for lead halide perovskites before, it is still not known what exact mechanism underlies these values. However, our present study can be used to argue that the observed strong FWM susceptibility is not due to resonant enhancement through one of the intermediate virtual states going ``on-shell''. Indeed, both the FWM channels $\Omega_1 = 2\omega_{\mathrm{NIR}}-\omega_{\mathrm{MIR}}$  and $\Omega_2 = \omega_{\mathrm{NIR}}+2\omega_{\mathrm{MIR}}$ remain comparable in intensity across the full spectral range despite the markedly different trajectories they follow on the energy diagram. Therefore, even though one might suspect a possible resonant trajectory for FWM1 ($\Delta < 2\,\hbar \omega_{\mathrm{NIR}}$), FMW2 always remains off-resonant, which suggests that FWM1 is off-resonant too. We conjecture that the lack of real on-shell intermediate level in FWM1 is a direct outcome of the specific orbital structure and matrix elements between conduction and valence bands in lead-halide perovskites \cite{PhysRevLett.130.106901,PhysRevB.107.125201}.
\vspace{0.2cm}

\noindent {\bf Conclusion}. We demonstrate tunable ultra-broadband frequency generation via four-wave mixing in lead–halide perovskites, which operates without the need for phase-matching, alignment, or dispersion engineering, and remains robust even for photon energies well below the bandgap. The central result of this work is that the conversion does not originate from the full bulk volume. Instead, it is localized near the crystal surfaces, where translational invariance is broken and phase-matching constraints are relaxed. In lead-halide perovskites, the exceptionally large intrinsic third-order nonlinearity makes this otherwise surface-limited interaction efficient enough to produce bright, collimated, and widely tunable emission. This highlights a fundamentally distinct regime of non-degenerate nonlinear response, in which surface-enabled frequency conversion and strong intrinsic third-order nonlinearity act in concert to produce broadband, stable emission.

This underscores the potential of these materials for compact, small-footprint photonic platforms.  Notably, lead-halide perovskites exhibit a high degree of compatibility with on-chip photonics. Numerous strategies for the micro- and nanoscale patterning of single-crystal perovskites have been demonstrated \cite{He2017, Lei2018, Zhang2020, Lei2020, ref1, Xu2023}, enabling their integration into compact photonic devices. The combination of large intrinsic nonlinearity, fabrication versatility, and alignment-insensitive operation distinguishes these materials from conventional nonlinear media and opens new opportunities for compact implementations of broadband frequency conversion.

These results establish lead–halide perovskites as a versatile platform for exploring third-order nonlinear optical phenomena in integrated settings, with immediate relevance for applications such as ultrafast pulse diagnostics, mid-infrared detection, and broadband optical waveform monitoring. More broadly, they point toward a new class of photonic architectures in which surface-driven nonlinear processes can be harnessed efficiently in scalable, small-footprint devices, advancing the development of next-generation nonlinear photonic technologies.
\vspace{2em}

\noindent{\large \textbf{Methods}}
\vspace{0.5em}

\noindent \textbf{Materials.} 
CH$_3$NH$_3$Br ($>$99.99\%) was purchased from GreatCell Solar. PbBr$_2$ (98\%), CsBr (99.9\% trace metal basis), DMF (anhydrous, 99.8\%), and DMSO (anhydrous, 99.9\%) were purchased from Sigma Aldrich. All chemicals were used as received without further purification.

\vspace{0.5cm}

\noindent \textbf{Growth of CH$_3$NH$_3$PbBr$_3$ Perovskite Single Crystals.} 
A 1.5\,M solution of CH$_3$NH$_3$Br/PbBr$_2$ in DMF was prepared, filtered through a 0.45\,$\mu$m-pore-size PTFE filter, and the vial containing 0.5--1\,ml of the solution was placed on a hot plate at 30\,$^\circ$C. The solution was then gradually heated to 60\,$^\circ$C and maintained at this temperature until the formation of CH$_3$NH$_3$PbBr$_3$ crystals. The crystals can be grown into larger sizes by elevating the temperature further. The crystals were collected and cleaned using a Kimwipe paper.

\vspace{0.5cm}

\noindent \textbf{Growth of CsPbBr$_3$ Perovskite Single Crystals.} 
A 1.0\,M solution of CsBr/2PbBr$_2$ in DMSO was prepared, and the vial containing 1.5\,ml of the solution was placed on a hot plate at 60\,$^\circ$C. The solution was gradually heated to 100\,$^\circ$C, during which yellow and orange crystals appeared as the temperature increased. Upon reaching 100\,$^\circ$C, the filtrate was transferred into a new vial preheated to 100\,$^\circ$C and maintained at this temperature for 3\,hours until the formation of CsPbBr$_3$ crystals. The crystals can be grown into larger sizes by elevating the temperature further. The crystals were collected, rinsed with hot DMSO, cleaned using a Kimwipe paper, and dried in a vacuum chamber at 100\,$^\circ$C for 1\,hour.

\noindent \textbf{Generation and detection of FWM.}
Tunable frequency mid-IR pulses are generated by an optical parametric amplifier (OPA; Light Conversion Orpheus-HE) pumped by a femtosecond laser system (Light Conversion Pharos) producing a train of pulses with a repetition rate of 3 kHz; central wavelength 1028 nm; pulse duration of 270fs and 2mJ/pulse. A small fraction (5\%) of the main beam from the amplifier is split off and used as a near-IR pulse, while the main part pumps the OPA to produce mid-IR pulses. Mid-IR and near-IR pulses are spatially and temporally focused on a 1\textit{mm}-thick, single crystalline CsPbBr$_3$ (and 2\textit{mm}-thick MAPbBr$_3$) sample using a $f=100$\textit{mm} ZnSe lens. Typical pulse energies were 0.5$\mu$J (near-IR) and 2.5$\mu$J (mid-IR). The generated FWM beam is spectrally analyzed at the sample exit using a fiber-coupled spectrometer (Ocean Optics Flame).

\noindent \textbf{Polarization dependent measurements.}
For the polarization dependent azimuthal scans, the power in the pump arm was first set by a pair of wire-grid polarizers (Thorlabs WP25H-Z). This is followed by a broadband tunable MIR quarter-wave plate (Alphalas PO-TWP-L4-25-IR) that turned the initial plane-polarized radiation into a circular polarization. Finally another wire-grid polarizer (Thorlabs WP25H-Z) was employed to rotate the polarization plane of the pump beam. The polarization plane in the probe arm was tuned by means of a GT polarizer (Thorlabs GT10) and a half-wave plate (Thorlabs WPH05M-1030).
\vspace{0.5em}

\noindent \textbf{Mid-IR-induced near-IR response.}
Mid-infrared pump and near-infrared probe pulses are focused onto a 1-mm-thick single-crystal CsPbBr$_3$ sample using an $f = 100$mm ZnSe lens. The mutual delay between the pulses $\tau$ is controlled by a mechanical delay stage in the near-IR arm.

\noindent \textit{Photodetector response.}
Measured by recording the total transmitted probe intensity as a function of delay using a Si photodiode detector (Thorlabs PDA100A2). The detector output is processed with a boxcar integrator (SRS SR250) and demodulated using a lock-in amplifier (SRS SR830) referenced to the pump modulation frequency.

\noindent \textit{NIR spectrum.}
It is obtained by spectrally filtering the transmitted probe with a monochromator (Horiba H10) and detecting the selected wavelength using an avalanche photodiode (Becker\&Hickl APM-400-P-078). By scanning the monochromator wavelength and repeating the delay-dependent measurement, the full two-dimensional signal is obtained.

\noindent \textbf{Author Contributions} A.S.K, D.L and Z.A. have performed the experiments; A.A.Z. and O.M.B. have synthesized the samples; A.S.K and Z.A. wrote the manuscript; Z.A. oversaw the project. The authors declare no competing interests.

\noindent \textbf{Acknowledgments.} We thank G. Koutentakis, R. Al Hyder and A. Volosniev for useful discussions. The work of DL was in part supported by APVV-23-0083.
\bibliography{mybib}

\end{document}